\documentclass[12pt]{article}
\usepackage{chetnew}
\usepackage{xparse}
\usepackage{amssymb, amsmath}
\usepackage{xcolor}
\usepackage{tikz}
\usepackage{cite}
\usepackage[linktocpage]{hyperref}
\usepackage{mathtools}

\DeclareFontShape{OT1}{cmr}{mx}{n}%
    {<->cmr10}{}

\renewcommand{\tilde}{\widetilde}

\renewcommand{\bar}{\overline}

\allowdisplaybreaks

\preprint{
NITEP 148}

\title{$S^1$ Reduction of 4D $\mathcal{N}=3$ SCFTs and \\[5mm] 
Squashing Independence of ABJM Theories}

\author{Tomoki Nakanishi$^{1,2\diamondsuit}$ and Takahiro
Nishinaka$^{1,2,3\clubsuit}$}
\affiliation{\smallskip  $^1$Department of Physics, Graduate School of Science\\
Osaka Metropolitan University, Osaka 558-8585, Japan\\
\bigskip
$^2$Nambu Yoichiro Institute of Theoretical and Experimental Physics (NITEP)\\
Osaka Metropolitan University, Osaka 558-8585, Japan\\
\bigskip
$^3$Osaka Central Advanced Mathematical Institute (OCAMI)\\
Osaka Metropolitan University, Osaka 558-8585, Japan
\emails{$^{\diamondsuit}$nakanishiphys@gmail.com, $^{\clubsuit}$nishinaka@omu.ac.jp}
}

\abstract{We study the compactification of 4D $\mathcal{N}=3$
superconformal field theories (SCFTs) on $S^1$, focusing on the relation 
between the 4D superconformal index and 3D partition function on the
squashed sphere $S^3_b$. Since the center $\mathfrak{u}(1)$ of the
$\mathfrak{u}(3)$ R-symmetry of the 4D theory can mix with an
$\mathcal{N}=6$ abelian flavor symmetry in three dimensions, the precise
4D/3D relation for the global symmetry is not obvious. Focusing on the case in which the 3D
theory is the ABJM theory we demonstrate that the above R-symmetry mixing can be
precisely identified by considering the Schur limit (and/or its
$\mathcal{N}=3$ cousin) of the 4D index. As a result, we generalize to
the ABJM theories recent discussions on the connection between
supersymmetry enhancement of the 4D index and squashing independence
of the $S^3_b$ partition function. 
 
}

\begin{document}

\maketitle
\toc

\section{Introduction}
\label{sec:intro}

There had been no known 3D $\mathcal{N}=6$ superconformal field theories
(SCFTs) until the authors of
\cite{Aharony:2008ug} discovered a special series of Chern-Simons matter
theories with this amount of supersymmetry. These theories are now
called ABJM theories, and are
identified as the low energy description of M2-branes on
$\mathbb{C}^4/\mathbb{Z}_k$.

Similarly, no 4D $\mathcal{N}=3$ SCFT had been known until the authors
of \cite{Garcia-Etxebarria:2015wns} discovered the first example using D3-branes on the
F-theory singularity $(\mathbb{C}^3\times T^2)/\mathbb{Z}_k$ for $k=3,4$
or $6$.\footnote{A general discussion on 4D $\mathcal{N}=3$
	SCFTs was
	first
	given in \cite{Aharony:2015oyb}. More detailed analysis on explicit
	constructions of 4D $\mathcal{N}=3$ SCFTs was given in
	\cite{Aharony:2016kai, Garcia-Etxebarria:2016erx}.} As discussed in \cite{Garcia-Etxebarria:2015wns}, the
duality between M-theory and F-theory suggests that the compactification
of these 4D $\mathcal{N}=3$ theories on $S^1$ gives rise to ABJM
theories (for specific values of the Chern-Simons level). Then it
would be desirable to find a clear correspondence between physical
quantities of 4D $\mathcal{N}=3$ SCFTs and ABJM theories.

One 4D quantity whose behavior under the compactification is
well-understood for generic $\mathcal{N}=2$ SCFTs is the superconformal index. Since the index can be thought
of as the (normalized) partition function of the 4D theory on $S^1 \times S^3_b$ with
background gauge fields turned on, it is reduced to the $S^3_b$
partition function of the 3D theory by the $S^1$ reduction. This
relation between the 4D index and $S^3_b$ partition function has been
carefully studied and confirmed in various examples \cite{Dolan:2011rp,
	Gadde:2011ia, Imamura:2011uw, Buican:2015hsa}.
This
leads us to comparing the superconformal index of the 4D $\mathcal{N}=3$
SCFTs and the $S^3_b$ partition function of the ABJM theories.

However, it turns out that the 3D reduction of
the R-symmetry of 4D $\mathcal{N}=3$ SCFTs is rather non-trivial, unlike
for generic 4D $\mathcal{N}=2$ SCFTs.
On the one hand, the bosonic (zero-form) global symmetry of the ABJM
theory is $\mathfrak{so}(6)_R\times \mathfrak{u}(1)_b$, where
$\mathfrak{u}(1)_b$ is a flavor symmetry commuting with the
$\mathcal{N}=6$ R-symmetry. On the other
hand, the bosonic (zero-form) global symmetry of 4D $\mathcal{N}=3$
SCFTs is $\mathfrak{u}(3)_R$
R-symmetry.\footnote{Here, we normalize the center $\mathfrak{u}(1)$ of
	$\mathfrak{u}(3)_R$ so that the 4D chiral supercharges $Q^I_\alpha$ are
	in the fundamental representation of $\mathfrak{u}(3)_R$. Note also that
	a 4D $\mathcal{N}=3$ SCFT cannot have a continuous
	$\mathcal{N}=3$ flavor symmetry without having extra supersymmetries \cite{Aharony:2015oyb}.
}  Naively, one might think that the
$\mathfrak{u}(3)_R$ in four dimensions is mapped to a $\mathfrak{u}(3)$ sub-algebra of
$\mathfrak{so}(6)_R$ under the compactification. However, this
naive map would lead to various
contradictions as discussed below.

First, the naive map $\mathfrak{u}(3)_R\hookrightarrow \mathfrak{so}(6)_R$ contradicts with a general relation between
supersymmetry enhancement of the 4D index and squashing
independence of the 3D partition function
\cite{Minahan:2021pfv}\footnote{Here,
we mean by ``supersymmetry enhancement'' that the number of preserved supercharges on
$S^1\times S^3_b$ is increased by taking a special limit of background
gauge fields.
}.
Here, by squashing
independence, we mean the property that the $b$-dependence of the
$S^3_b$ partition function $Z_{S^3_b}$ can be absorbed by rescaling mass
parameters. Such a phenomenon
was recently observed in \cite{Chester:2021gdw} for the ABJM theory when a special
value of an (imaginary) mass parameter is turned on. This was then interpreted in \cite{Minahan:2021pfv} as a
supersymmetry enhancement on $S^3_b$, or on $S^1\times S^3_b$ in the
case when the
theory has a 4D uplift. In the latter case, the special value of the 3D mass
parameter corresponds to a limit of a fugacity in the 4D superconformal
index.
In particular, the Schur limit of the 4D
index (i.e., Schur index) generally leads to squashing independence of $Z_{S^3_b}$
under the compactification \cite{Minahan:2021pfv}. As we will show in Sec.~\ref{sec:mixing}, the naive map $\mathfrak{u}(3)_R
	\hookrightarrow \mathfrak{so}(6)_R$ discussed above, however, contradicts with this general relation
between the 4D Schur
limit and 3D squashing independence.

Another contradiction of the naive map
$\mathfrak{u}(3)_R\hookrightarrow \mathfrak{so}(6)_R$ can be seen in the moduli space of vacua. Let us consider the case of
rank-one $\mathcal{N}=3$ SCFTs \cite{Nishinaka:2016hbw}. The
moduli space is then $\mathbb{C}^3/\mathbb{Z}_k$ for $k=3,4$ or $6$, where the $\mathbb{Z}_k$-action is such that
$\mathbb{C}^3\ni (z_1,z_2,z_3)\to (\omega
	z_1,\,\omega^{-1}z_2,\,\omega z_3)$ for $\omega \in
	\mathbb{Z}_k$. The 3D reduction of this 4D $\mathcal{N}=3$ SCFT is
believed to be the ABJM theory with gauge group $U(1)_k\times
	U(1)_{-k}$, whose moduli space is $\mathbb{C}^4/\mathbb{Z}_k$
with the $\mathbb{Z}_k$-action such that
$\mathbb{C}^4\ni (z_1,z_2,z_3,z_4) \to (\omega z_1,\,\omega^{-1}z_2,\,\omega
	z_3,\,\omega^{-1}z_4)$. The emergence of the extra direction (parameterized by $z_4$)
is a general phenomenon for the compactification of 4D $\mathcal{N}\geq 2$
theories; the dimension of the Coulomb branch is doubled by the
compactification. Here, $z_1,z_2$ and $z_3$ are charged under the 4D $\mathfrak{u}(3)_R$ symmetry while $z_4$ is neutral.\footnote{We here assume there is no non-trivial quantum corrections
	to the moduli space under the $S^1$ compactification, which we believe
	is a
	natural assumption with this large amount of supersymmetry.} From the 3D
viewpoint, however, the $\mathfrak{so}(6)_R$ R-symmetry non-trivially acts on all
$z_1,z_2,z_3$ and $z_4$.\footnote{To be more specific, $(z_1,\bar{z}_2,z_3)$ forms a
	fundamental representation of
	$\tilde{\mathfrak{u}(3)}_R$, while $(z_1,\bar{z}_2, z_3,\bar{z}_4)$ forms a
	fundamental representation of $\mathfrak{su}(4)\simeq
		\mathfrak{so}(6)_R$. Here, the action of $\widetilde{\mathfrak{u}(3)}_R$ is different
	from that of $\mathfrak{u}(3)_R$ only in the action of its center; the former is obtained by
	rescaling the charge of
	$\mathfrak{u}(1)_c\subset \mathfrak{u}(3)_R$ by $1/2$.
	Indeed,
	our normalization of the
	charge of $\mathfrak{u}(1)_c\subset \mathfrak{u}(3)_R$ is such that the
	4D chiral supercharges
	$\mathcal{Q}^I_\alpha$ are in its fundamental representation, while the
	action of the center of $\widetilde{\mathfrak{u}(3)}_R$ is such that
	$(z_1,\bar{z}_2,z_3)$ are in its fundamental
	representation. Since $z_1,z_2$ and $z_3$ are identified as the VEVs of the
	three chiral multiplets in an $\mathcal{N}=3$ vector multiplet (which is
	equivalent to an $\mathcal{N}=4$ vector multiplet for CPT reasons), we see
	that these normalizations differ by a factor two. The authors thank
	Y.~Tachikawa for pointing out this fact.
} Therefore, if the naive inclusion
$\mathfrak{u}(3)_R\hookrightarrow \mathfrak{so}(6)_R$ is correct, there must be a $\mathfrak{u}(3)$
sub-algebra of $\mathfrak{so}(6)_R$ that keeps $z_4$ invariant. It turns
out, however, that there is no
such sub-algebra.

In this paper, we show that the above two contradictions are resolved
when the $\mathfrak{u}(3)_R$ symmetry in four dimensions is mixed with the
$\mathfrak{u}(1)_b$ symmetry in three dimensions under the $S^1$ compactification.
Indeed, we show that there is a unique
$\mathfrak{u}(3)$ sub-algebra of $\mathfrak{so}(6)_R\times
	\mathfrak{u}(1)_b$ that, when identified as the 3D reduction of the 4D
$\mathfrak{u}(3)_R$ symmetry, resolves all the above
contradictions. Our result would be important when comparing the
superconformal index of the 4D $\mathcal{N}=3$ theories and the sphere
partition function of their 3D reductions. Indeed, given the above mixing between $\mathfrak{u}(3)_R$ and
$\mathfrak{u}(1)_b$, one can now establish a clear correspondence
between the supersymmetry enhancement of the 4D $\mathcal{N}=3$ SCFT on
$S^1\times S^3_b$ and the squashing
independence of the ABJM theories on $S^3_b$.

The organization of this paper is as follows.
In Sec.~\ref{sec:squashing}, we give a brief review of the relation
between the supersymmetry enhancement of the 4D index and the squashing
independence of the 3D partition function. We also discuss its
$\mathcal{N}=3$ cousin there. In Sec.~\ref{sec:mixing}, we consider the
$S^1$-compactification of the 4D $\mathcal{N}=3$ SCFT and discuss the
R-symmetry mixing, focusing on the case that the 3D theory is the ABJM
theory. In Sec.~\ref{sec:divergence}, we discuss a divergence that
arises in the reduction of 4D $\mathcal{N}=3$ superconformal index. We
finally conclude in Sec.~\ref{sec:conclusions}.
In appendix
\ref{app:R-symmetry}, we discuss in some more detail how the 4D
and 3D symmetries are related under the $S^1$-compactification.

\section{Squashing independence of 3D theories arising from 4D
  $\mathcal{N}\geq 2$ SCFTs}

\label{sec:squashing}

\begin{table}
	\begin{center}
		\begin{tabular}{|c|c|c|c|c|c|c|c|}
			\hline
			                                 & $j_1$           & $j_2$           & $R$            & $r$            & $f$  & $j_2+j_1-r$ & $j_2-j_1-R-2r-\frac{3}{2}f$ \\
			\hline
			\hline
			$\mathcal{Q}^1{}_+$              & $+ \frac{1}{2}$ & $0$             & $+\frac{1}{2}$ &
			$+\frac{1}{2}$                   & $0$             & $0$             & $-2$                                                                               \\
			$\mathcal{Q}^1{}_-$              & $- \frac{1}{2}$ & $0$             & $+\frac{1}{2}$ &
			$+\frac{1}{2}$                   & $0$             & $-1$            & $-1$                                                                               \\
			$\mathcal{Q}^2{}_+$              & $+ \frac{1}{2}$ & $0$             & $-\frac{1}{2}$ &
			$+\frac{1}{2}$                   & $0$             & $0$             & $-1$                                                                               \\
			$\mathcal{Q}^2{}_-$              & $- \frac{1}{2}$ & $0$             & $-\frac{1}{2}$ &
			$+\frac{1}{2}$                   & $0$             & $-1$            & $0$                                                                                \\
			$\tilde{\mathcal{Q}}_{1\dot{+}}$ & $0$             & $+\frac{1 }{2}$ &
			$-\frac{1}{2}$                   & $-\frac{1}{2}$  & $0$             & $1$            & $2$                                                               \\
			$\tilde{\mathcal{Q}}_{1\dot{-}}$ & $0$             & $-\frac{1 }{2}$ &
			$-\frac{1}{2}$                   & $-\frac{1}{2}$  & $0$             & $0$            & $1$                                                               \\
			$\tilde{\mathcal{Q}}_{2\dot{+}}$ & $0$             & $+ \frac{1}{2}$ &
			$+\frac{1}{2}$                   & $-\frac{1}{2}$  & $0$             & $1$            & $1$                                                               \\
			$\tilde{\mathcal{Q}}_{2\dot{-}}$ & $0$             & $- \frac{1}{2}$ &
			$+\frac{1}{2}$                   & $-\frac{1}{2}$  & $0$             & $0$            & $0$                                                               \\
			\hline
			$\mathcal{Q}^3{}_+$              & $+ \frac{1}{2}$ & $0$             & $0$            & $-\frac{1}{2}$ & $+1$
			                                 & $1$             & $-1$                                                                                                 \\
			$\mathcal{Q}^3{}_-$              & $- \frac{1}{2}$ & $0$             & $0$            & $-\frac{1}{2}$ & $+1$
			                                 & $0$             & $0$                                                                                                  \\
			$\tilde{\mathcal{Q}}_{3\dot{+}}$ & $0$             & $+ \frac{1}{2}$ & $0$            &
			$+\frac{1}{2}$                   & $-1$            & $0$             & $1$                                                                                \\
			$\tilde{\mathcal{Q}}_{3\dot{-}}$ & $0$             & $- \frac{1}{2}$ & $0$            &
			$+\frac{1}{2}$                   & $-1$            & $-1$            & $0$                                                                                \\
			\hline
		\end{tabular}
		\caption{Quantum numbers of supercharges}
		\label{table:supercharge}
	\end{center}
\end{table}

\subsection{4D $\mathcal{N}=2$ SCFTs and 3D $\mathcal{N}=4$ theories}

\label{subsec:review}

Here, we briefly review the relation between limits of 4D superconformal
index and the 3D squashing independence \cite{Minahan:2021pfv}.

The superconformal index of a 4D $\mathcal{N}=2$ SCFT is defined by
\begin{align}
	I = \text{Tr}(-1)^Fp^{j_2-j_1-r}q^{j_2+j_1-r}t^{r+R}\prod_{i}a_i^{f_i}~,
	\label{eq:index-1}
\end{align}
where the trace is taken over the space of local operators, $(j_1,j_2)$ are $\mathfrak{so}(4)$ spins, $R, r$ are respectively
the $SU(2)_R$ and $U(1)_r$ charges, and $f_i$ are the flavor charges. In
terms of the generators of the 4D $\mathcal{N}=2$ superconformal algebra
reviewed in appendix \ref{app:4Dsca}, $R$ and $r$ are written as
\begin{align}
	R \equiv \frac{1}{2}\left(\mathcal{R}^1{}_1 -
	\mathcal{R}^2{}_2\right)~,\qquad r\equiv \mathcal{R}^1{}_1 +
	\mathcal{R}^2{}_2~.
\end{align}
Since the exponents of $p,q,t$ and $a_i$ commute with
$\tilde{\mathcal{Q}}_{2\dot-}$ and $\tilde{\mathcal{S}}^{2\dot-}$ (Table \ref{table:supercharge}), the index receives contributions only from
operators annihilated by these supercharges. Therefore we say that the
above index preserves two supercharges.

The superconformal index has several interesting limits of
parameters. In particular, the Schur limit, $t\to q$, is known to
provide a variety of application.\footnote{For instance, it gives rise to a general connection between 4D $\mathcal{N}=2$ SCFTs and
	non-unitary vertex algebras \cite{Beem:2013sza}. It also simplifies the TQFT
	expression for the
	index of class $\mathcal{S}$ theories \cite{Gadde:2011ik}.} One important feature
of the Schur limit is that
the resulting index (called the ``Schur index'') turns out to be
independent of $p$ \cite{Gadde:2011uv, Minahan:2021pfv}. The reason for this is the following.  First note
that, when $t=q$,
the index is written as
\begin{align}
	I= \text{Tr}(-1)^Fp^{j_2-j_1-r}q^{j_2+j_1+R}\prod_{i}a_i^{f_i}~.
\end{align}
There are now four supercharges, $\mathcal{Q}^1{}_-,\,\tilde{\mathcal{Q}}_{2\dot{-}},\,\mathcal{S}_1{}^-$ and
$\tilde{\mathcal{S}}^{2\dot{-}}$, commuting with the exponents of $p,q$ and
$a_i$. Therefore, the Schur index preserves four
supercharges, and receives contributions only from operators
annihilated by them.
We see that the exponent of $p$,
\begin{align}
	j_2 - j_1-r = \left\{\mathcal{Q}^1{}_-, \mathcal{S}_1{}^-\right\} - \left\{\tilde{\mathcal{Q}}_{2\dot{-}},\tilde{\mathcal{S}}^{2\dot{-}}\right\}~,
\end{align}
vanishes for all such operators.
As a result, the Schur index has no $p$-dependence and is expressed as
\begin{align}
	I\big|_{t\to q} = \text{Tr}(-1)^F q^{j_2+j_1+R}\prod_{i}a_i^{f_i}~.
	\label{eq:Schur2}
\end{align}
The absence of $p$-dependence is clearly a consequence of the
supersymmetry enhancement triggered by $t\to q$ \cite{Gadde:2011uv, Minahan:2021pfv}.

Let us now consider the compactification of the 4D $\mathcal{N}=2$
SCFT on $S^1$. In the deep infrared, we end up with a 3D $\mathcal{N}=4$ SCFT. Its
$S^3_b$ partition function can be obtained as the small $S^1$ limit of the 4D
superconformal index \eqref{eq:index-1} \cite{Dolan:2011rp,
	Gadde:2011ia, Imamura:2011uw, Buican:2015hsa, Bullimore:2014nla}. In this limit,  we
set
\begin{align}
	p=e^{-\beta b}~,\qquad q= e^{-\beta b^{-1}}~,\qquad \frac{t}{\sqrt{pq}} =
	e^{-i\beta m}~,\qquad a_i = e^{-i\beta M_i}~,
	\label{eq:fugacity-relation}
\end{align}
and take the limit $\beta \to 0$ with $b,m$ and $M_i$ kept fixed.\footnote{The
	special combination $t/\sqrt{pq}$ appears here since it is a flavor fugacity from the viewpoint of the $\mathcal{N}=1$
	supersymmetry involving the preserved supercharge $\tilde{Q}_{2\dot-}$.} Then
the 4D index reduces to the partition function $Z_{S^3_b}$ of the 3D $\mathcal{N}=4$
theory on $S^3_b$ (up to a prefactor). The parameters $m$ and $M_i$ are
identified as mass deformation parameters of the 3D theory.

From the 3D viewpoint, the Schur limit $t\to q$ corresponds to constraining
the 3D mass parameter $m$ as
\begin{align}
	m = i\frac{b-b^{-1}}{2}~.
	\label{eq:Schur}
\end{align}
With this value of $m$, the supersymmetry preserved on $S^3_b$
is enhanced. As a result, the $b$-dependence of $Z_{S^3_b}$
can be removed by rescaling the remaining mass parameters $M_i$
\cite{Minahan:2021pfv}. Indeed, the index is now independent of
$p$, and therefore the $b$-dependence is only in $q =
	e^{-\beta b^{-1}}$. We see that this $b$-dependence can be removed by
\begin{align}
	\beta \to \beta b~,\qquad M_i\to b^{-1}M_i~,
\end{align}
where rescaling $M_i$ is necessary for keeping $a_i= e^{-i\beta M_i}$ unchanged.
Hence, the Schur limit $t\to q$ in four dimensions generally leads to
squashing independence in three dimensions \cite{Minahan:2021pfv}.

It was also found in \cite{Minahan:2021pfv} that the 3D squashing independence also occurs at
\begin{align}
	m =
	-i\frac{b+b^{-1}}{2}~.
	\label{eq:Coulomb-m}
\end{align}
In four dimensions, this corresponds to the condition $t=pq$, under which the 4D index
is expressed as
\begin{align}
	I\big|_{t=pq} = \text{Tr}(-1)^F
	p^{j_2-j_1+R}q^{j_2+j_1+R}\prod_{i}a_i^{f_i}~.
\end{align}
This index counts operators annihilated by
$\tilde{\mathcal{Q}}_{1\dot+},\,\tilde{\mathcal{Q}}_{2\dot-},\,\tilde{\mathcal{S}}^{1\dot+}$ and
$\tilde{\mathcal{S}}^{2\dot-}$~, and for such operators $j_2 + R =
	\frac{1}{2}\{\tilde{\mathcal{Q}}_{1\dot+},\,\tilde{\mathcal{S}}^{1\dot+}\} -
	\frac{1}{2}\{\tilde{\mathcal{Q}}_{2\dot-},\,\tilde{\mathcal{S}}^{2\dot-}\}$
vanishes. Therefore, the above expression for the index reduces to
\begin{align}
	I\big|_{t=pq} = \text{Tr}(-1)^F
	\left(\frac{q}{p}\right)^{j_1}\prod_{i}a_i^{f_i}~.
	\label{eq:Coulomb}
\end{align}
Since the index involves only one combination of the superconformal
fugacities, $q/p$, its 3D reduction leads to $Z_{S^3_b}$ whose
$b$-dependence can again be removed by rescaling mass parameters. Note,
however, that the index \eqref{eq:Coulomb} is always divergent when the 4D
theory has a freely generated Coulomb branch, since all the Coulomb branch operators
contribute to it with $j_1=R=f_i=0$.\footnote{By ``Coulomb branch
	operators,'' we mean local operators annihilated by all anti-chiral
	$\mathcal{N}=2$ supercharges. For the fact that $f_i=0$
	for all Coulomb branch operators, see \cite{Buican:2013ica, Buican:2014qla}.}

\subsection{4D $\mathcal{N}=3$ SCFTs and 3D $\mathcal{N}=6$ theories}
\label{subsec:N=3}

When the 4D theory has an extra supersymmetry, one can generalize the
above discussion. Such a generalization will be useful when we discuss
in the next section the relation between 4D $\mathcal{N}=3$ SCFTs and 3D
ABJM theories.

We will focus on 4D $\mathcal{N}=3$ SCFTs without $\mathcal{N}=4$
supersymmetry, although its generalization to 4D $\mathcal{N}=4$ SCFTs is
straightforward and will be discussed in \cite{Nakanishi:future}.
The bosonic global symmetry of 4D (genuine) $\mathcal{N}=3$ SCFT is just the $\mathfrak{u}(3)_R$ symmetry;
there is no ``$\mathcal{N}=3$ flavor symmetry.'' This
$\mathfrak{u}(3)_R$ contains the $\mathcal{N}=2$ R-symmetry, $\mathfrak{su}(2)_R\times
	\mathfrak{u}(1)_r$, and the $\mathcal{N}=2$ flavor symmetry,
$\mathfrak{u}(1)_f$. Therefore, the most general expression for the superconformal index is
written as
\begin{align}
	I = \text{Tr}(-1)^Fp^{j_2-j_1-r}q^{j_2+j_1-r}t^{r+R}a^{f}~,
	\label{eq:index0}
\end{align}
where we define the $\mathfrak{u}(1)_f$ charge by
\begin{align}
	f\equiv \mathcal{R}^1{}_1+\mathcal{R}^2{}_2 +
	2\mathcal{R}^3{}_3~,
	\label{eq:f}
\end{align}
as in \cite{Nishinaka:2016hbw}.
For generic values of the fugacities, the above index preserves only
$\tilde{\mathcal{Q}}_{2\dot-}$ and $\tilde{\mathcal{S}}^{2\dot-}$.
Since the discussions in the previous sub-section are also applied to
$\mathcal{N}=3$ SCFTs, we see that there are two special values of the
fugacity $t$ that lead to supersymmetry enhancement:
\begin{align}
	t = q~,\qquad t = pq~.
\end{align}
In three dimensions, they correspond to two special values of
a mass parameter that give rise to squashing independence.

We now ask whether a similar supersymmetry enhancement occurs at a special
value of the fugacity $a$. If the theory has only $\mathcal{N}=2$
supersymmetry with a flavor charge $f$, no
such supersymmetry enhancement occurs since $f$ commutes with
all $\mathcal{N}=2$ supercharges. However, when the theory has an $\mathcal{N}=3$
supersymmetry and $f$ is defined by \eqref{eq:f}, there exists a supercharge that does {\it not} commute with
$f$. Such a supercharge can always be preserved by setting $a$ to a
special value.

For instance, one can
preserve $\mathcal{Q}^3{}_-$ (and its conjugate $\mathcal{S}_3{}^-$) by setting
\begin{align}
	a = \frac{\sqrt{t}}{p}~.
	\label{eq:cond1}
\end{align}
In this case, the expression for the index reduces to
\begin{align}
	I\big|_{a=\frac{\sqrt{t}}{p}} = \text{Tr}(-1)^F p^{j_2-j_1-r-f}q^{j_2+j_1-r}t^{r+R+\frac{f}{2}}~.
	\label{eq:index1}
\end{align}
We see from Table \ref{table:supercharge} that it indeed preserves
$\mathcal{Q}^3{}_-$ and $\mathcal{S}_3{}^-$ in addition to
$\tilde{\mathcal{Q}}_{2\dot-}$ and $\tilde{\mathcal{S}}^{2\dot-}$. Note
that one can rewrite  \eqref{eq:index1} as
\begin{align}
	I\big|_{a=\frac{\sqrt{t}}{p}} = \text{Tr}(-1)^F
	\left(pt\right)^{\frac{1}{2}  \left(-\delta_1 +
	\delta_2\right)} q^{j_2+j_1-r}
	\left(\frac{p}{t}\right)^{\frac{1}{2}(j_2-j_1-R-2r-\frac{3}{2}f)}
	\label{eq:pre-index}
\end{align}
where $\delta_1 \equiv \{\tilde{\mathcal{Q}}_{2\dot -},\, \tilde{\mathcal{S}}^{2\dot-}\}
	\frac{1}{2}E -j_2 - R + \frac{1}{2}r$ and  $\delta_2 \equiv \{ \mathcal{Q}^3{}_-,\, \mathcal{S}_3{}^-\} =
	\frac{1}{2}E - j_1 - \frac{1}{2}f+\frac{1}{2}r$.
Since the index is contributed only from operators annihilated by
$\tilde{\mathcal{Q}}_{2\dot
		-},\,\mathcal{Q}^3{}_-,\,\tilde{\mathcal{S}}^{2\dot-}$ and $\mathcal{S}_3{}^-$,
the expression \eqref{eq:pre-index} can be reduced to
\begin{align}
	I\big|_{a=\frac{\sqrt{t}}{p}} = \text{Tr}(-1)^F
	q^{j_2+j_1-r}\left(\frac{p}{t}\right)^{\frac{1}{2}\left(j_2-j_1-R-2r-\frac{3}{2}f\right)}~.
	\label{eq:index2}
\end{align}

The fact that the $(pt)$-dependence drops out in \eqref{eq:index2}
implies a new condition for the squashing independence in three dimensions. To see this,
first let us write things in terms of the 3D parameters using
\begin{align}
	p=e^{-\beta b}~,\qquad q= e^{-\beta b^{-1}}~,\qquad \frac{t}{\sqrt{pq}} =
	e^{-i\beta m}~,\qquad a = e^{-i\beta M}~.
	\label{eq:mM}
\end{align}
The condition \eqref{eq:cond1} is then translated as
\begin{align}
	M = \frac{m}{2}+i\frac{3b-b^{-1}}{4}~.
	\label{eq:M}
\end{align}
Under this condition, the index depends only on $q= e^{-\beta b^{-1}}$
and $p/t = e^{i\beta m - \beta \frac{b-b^{-1}}{2}}$, and therefore the $b$-dependence can be removed
by the replacement
\begin{align}
	\beta \to \beta b~,\qquad m \to b^{-1}m-i\frac{b-b^{-1}}{2}~.
\end{align}
This implies that, when \eqref{eq:M} is imposed, the $b$-dependence of
the $S^3_b$ partition
function can be removed by $m \to b^{-1}m-i\frac{b-b^{-1}}{2}$.

\section{R-symmetry mixing under the $S^1$-compactification}
\label{sec:mixing}

We here demonstrate that the above general discussion on squashing
independence of 3D $\mathcal{N}\geq 4$ theories puts a strong constraint
on the relation between the R-symmetries of a 4D $\mathcal{N}=3$
SCFT and its 3D reduction. For simplicity, we focus on 4D SCFTs whose 3D reduction is the ABJM
theory. Since the bosonic (zero-form) global symmetry is enhanced from
$\mathfrak{u}(3)_R$ to $\mathfrak{so}(6)_R \times \mathfrak{u}(1)_b$ by
the compactification, we need to understand which $\mathfrak{u}(3)$
sub-algebra of the latter corresponds to the 3D reduction of the
former. While a naive expectation might be that $\mathfrak{u}(3)_R$
simply descends to a sub-algebra of $\mathfrak{so}(6)_R$, that would
contradicts with the general discussions in the previous section.

\subsection{4D fugacities and 3D masses}
\label{subsec:4D3D}

To see the above statement, let us first make a concrete connection
between the R-charges in four and three dimensions.

We denote the 3D supercharges by $\mathsf{Q}^s_\alpha$ for
$s=1,\cdots,6$ so that we can
distinguish them from the 4D supercharges $\mathcal{Q}^I{}_\alpha$ and
$\tilde{\mathcal{Q}}_{I\dot\alpha}$. We arrange them so that
$\mathcal{Q}^I{}_\alpha$ and $\tilde{\mathcal{Q}}_{I\dot \alpha}$ descend to
$\frac{1}{\sqrt{2}}(\mathsf{Q}^{2I-1}{}_\alpha\pm
	i\mathsf{Q}^{2I}{}_{\alpha})$ in three dimensions.
We also denote by
$R^\text{4D}_I$ and $R^\text{3D}_I$ a basis of $\mathfrak{u}(3)_R$ and
$\mathfrak{so}(6)_R$, respectively. We take them so that
$\mathcal{Q}^I{}_\alpha$ and $\tilde{\mathcal{Q}}_{I\dot\alpha}$
respectively have charge
$+1$ and $-1$ under $R^{\text{4D}}_I$ but are neutral under $R^\text{4D}_{J(\neq I)}$, and
similarly that $\mathsf{Q}^{2I-1}{}_\alpha\pm i\mathsf{Q}^{2I}{}_\alpha$ have charge $\pm1$ under $R^\text{3D}_I$
but are neutral under $R^\text{3D}_{J(\neq I)}$. Note that, if the 4D R-symmetry
$\mathfrak{u}(3)_R$ was mapped into $\mathfrak{so}(6)_R$ by the compactification, we
would have
\begin{align}
	R_I^\text{4D} = R_I^\text{3D}~.
	\label{eq:wrong}
\end{align}
However, since the center $\mathfrak{u}(1)_c\subset \mathfrak{u}(3)_R$
can be mixed with $\mathfrak{u}(1)_b$ in the ABJM theory, the most
general identification is written as
\begin{align}
	R^\text{4D}_I =  R^\text{3D}_I + \xi J^\text{3D}_{U(1)_b}~,
	\label{eq:4D-3D}
\end{align}
where $J^\text{3D}_{U(1)_b}$ is the charge of the $\mathfrak{u}(1)_b$ flavor
symmetry in three dimensions, and $\xi$ is a real number.\footnote{As
	discussed below, we will normalize $J^\text{3D}_{U(1)_b}$ so that the
	3D twisted chiral multiplet with $R^\text{3D}_1 = R^\text{3D}_3= 1/2$ has charge $J^\text{3D}_{U(1)_b} =
		1/2$.} When $\xi\neq 0$, there is a mixing between $\mathfrak{u}(1)_c$ and $\mathfrak{u}(1)_b$.
Indeed, we will show below that $\xi$ does not vanish.

Let us now write the superconformal index \eqref{eq:index0} of
the 4D theory in terms of $R_i^\text{3D}$ and $J^\text{3D}_{U(1)_b}$. From Table
\ref{table:supercharge}, we see that
\begin{align}
	R^\text{4D}_1 = R+r+\frac{f}{2}~,\qquad R^\text{4D}_2 = -R + r +
	\frac{f}{2}~,\qquad R^\text{4D}_3 = f~.
\end{align}
This implies that the index \eqref{eq:index0} can be expressed as
\begin{align}
	I = \text{Tr}(-1)^F p^{j_2-j_1} q^{j_2+j_1}
	\left(\frac{t}{\sqrt{pq}}\right)^{R^\text{4D}_1}\left(\frac{1}{\sqrt{pq}}\right)^{R^\text{4D}_2}\left(a\sqrt{\frac{pq}{t}}\right)^{R^\text{4D}_3}~.
\end{align}
Using the identification \eqref{eq:4D-3D}, one can rewrite this in
terms of the 3D R-charges and $J_{U(1)_b}^\text{3D}$ as
\begin{align}
	I = \text{Tr}(-1)^F p^{j_2-j_1} q^{j_2+j_1}
	\left(\frac{t}{\sqrt{pq}}\right)^{R^\text{3D}_1}\left(\frac{1}{\sqrt{pq}}\right)^{R^\text{3D}_2}\left(a\sqrt{\frac{pq}{t}}\right)^{R^\text{3D}_3}\left(a
	\sqrt{\frac{t}{pq}}\right)^{\xi
	J^\text{3D}_{U(1)_b}}~.
	\label{eq:index11}
\end{align}
Since this index preserves $\tilde{Q}_{2\dot-}$, one can regard it as an
$\mathcal{N}=1$ index associated with
$(Q^2{}_\alpha,\,\tilde{Q}_{2\dot\alpha})$. In this case,
$t/\sqrt{pq}$ and $a\sqrt{pq/t}$ are regarded as {\it flavor} fugacities from the
$\mathcal{N}=1$ viewpoint. In three dimensions, these fugacities reduce to two mass parameters
associated with a 3D $\mathcal{N}=2$ flavor symmetry. For the ABJM theory, the $\mathcal{N}=2$ flavor symmetry is
the product of $\mathfrak{so}(4)\subset \mathfrak{so}(6)_R$ and
$\mathfrak{u}(1)_b$. While this product is of
rank three, only two independent mass parameters arise in the
reduction of the 4D index, since the remaining one is associated with an
accidental symmetry in three dimensions.

To be more specific, let us denote by $m_2$ and $m_3$ the mass parameters associated with
$\mathfrak{so}(2)^2\subset \mathfrak{so}(4)\subset \mathfrak{so}(6)_R$, and by $m_1$ that
associated with $\mathfrak{u}(1)_b$, as in \cite{Chester:2021gdw}.
The expression
\eqref{eq:index11} implies that these masses are related to the 4D
fugacities as\footnote{Note here that $R_2^\text{3D}$ is identified as
	the superconformal R-charge of the 3D $\mathcal{N}=2$ superconformal
	algebra associated with $\mathsf{Q}^3_\alpha \pm i\mathsf{Q}^4_\alpha$. Therefore, $(1/\sqrt{pq})^{R^\text{3D}_2}$ in \eqref{eq:index11}
	corresponds to the background gauge field for the R-symmetry that reduces
	to the 3D $\mathcal{N}=2$ superconformal
	R-symmetry under the compactification. Turning on such a background gauge field is
	necessary for preserving the supersymmetry on $S^1\times S^3$ \cite{Festuccia:2011ws}.}
\begin{align}
	\frac{t}{\sqrt{pq}} = e^{-i\beta m_2}~,\qquad a\sqrt{\frac{pq}{t}} =
	e^{-i\beta m_3}~,\qquad \left(a\sqrt{\frac{t}{pq}}\right)^\xi = e^{-i\beta m_1}~.
\end{align}
Since $p$ and $q$ are related to the squashing parameter by
$p=e^{-\beta b}$ and $q=e^{-\beta b^{-1}}$~, for fixed $b$ only two of
$m_1,m_2$ and $m_3$ are independent.
In terms of $m$ and $M$ defined in \eqref{eq:mM}, the mass parameters
$m_i$ are expressed as
\begin{align}
	m_1 = \xi\left(M + \frac{m}{2} + i\frac{b+b^{-1}}{4}\right)~,\qquad m_2
	= m~,\qquad m_3 = M -\frac{m}{2}- i\frac{b+b^{-1}}{4}
	\label{eq:dict}
\end{align}

We stress that the three mass parameters of the ABJM theory are
constrained here as above since one linear combination of them is associated
with an accidental symmetry in three dimensions and does not have a 4D
counterpart. One important question is then which two linear
combinations have 4D counterparts. The answer to this question depends on
the value of $\xi$, which
characterizes the mixing between $\mathfrak{u}(1)_c$ and
$\mathfrak{u}(1)_b$. Below, we will uniquely fix the value of $\xi$ so
that it is consistent with the general discussion on squashing independence in the
previous section.

\subsection{R-symmetry mixing consistent with the squashing independence}

Let us consider the Schur limit of the 4D index by imposing
\eqref{eq:Schur}. Under \eqref{eq:dict}, this condition is equivalent to
\begin{align}
	m_1 = \xi\left(M + i\frac{b}{2}\right)~,\qquad m_2 =
	i\frac{b-b^{-1}}{2}~,\qquad m_3 = M- i\frac{b}{2}
	\label{eq:cond-Schur}
\end{align}
Recall that the Schur limit of the 4D index generally leads to $Z_{S^3_b}$
whose $b$-dependence can be removed by rescaling the remaining mass
parameter $M$ as
\begin{align}
	M \to b^{-1}M~.
	\label{eq:rescale-Schur}
\end{align}
Here we will show that this general constraint uniquely fixes the value
of $\xi$ in \eqref{eq:cond-Schur}.

To that end, let us first look at $Z_{S^3_b}$ of the ABJM theory. The supersymmetric localization
leads to the following formula for $Z_{S^3_b}$ \cite{Kapustin:2009kz, Hama:2010av, Hama:2011ea}:
\begin{align}
	Z_{S^3_b} & \sim \int\frac{\prod_{i=1}^Nd\mu_i d\nu_i}{(N!)^2}e^{i\pi k\left(\sum_{i=1}^N(\nu_i^2
	- \mu_i^2)\right)}\prod_{i<j}\sinh\left(\pi
	b(\mu_i-\mu_j)\right)\sinh\left(\pi b^{-1}(\mu_i-\mu_j)\right)
	\nonumber                                                                                         \\
	          & \qquad \times \prod_{i<j}\sinh \left(\pi
	b(\nu_i-\nu_j)\right)\sinh\left(\pi b^{-1}(\nu_i-\nu_j)\right)
	\nonumber                                                                                         \\
	          & \qquad \times\prod_{i,j=1}^N\Bigg[
		s_b\left(\frac{iQ}{4} - \left(\mu_i-\nu_j +
		\frac{m_1+m_2+m_3}{2}\right)\right)s_b\left(\frac{iQ}{4}
		-\left(\mu_i-\nu_j + \frac{m_1-m_2-m_3}{2}\right)\right)
	\nonumber                                                                                         \\
	          & \qquad \times s_b\left(\frac{iQ}{4} -\left(-\mu_i+\nu_j +
		\frac{-m_1-m_2+m_3}{2}\right)\right)s_b\left(\frac{iQ}{4} -
		\left(-\mu_i+\nu_j + \frac{-m_1+m_2-m_3}{2}\right)\right)
		\Bigg]~,
	\label{eq:ZS3}
\end{align}
where $s_b(z)$ is the double-sine function
\begin{align}
	s_b(z) \equiv \prod_{k,\ell=0}^\infty\frac{\left(k+\frac{1}{2}\right)b +\left(\ell+\frac{1}{2}\right)b^{-1} -iz}{\left(k+\frac{1}{2}\right)b + \left(\ell+\frac{1}{2}\right)b^{-1}+iz}~.
\end{align}
Recall here that $m_1$ is the mass parameter associated with $\mathfrak{u}(1)_b$,
and $m_2$ and $m_3$ are those associated with $\mathfrak{so}(2)^2
	\subset \mathfrak{so}(4)$. The above normalizations of $m_1,m_2$ and
$m_3$ imply that a 3D (twisted) chiral multiplet has $R_1^\text{3D} =
	R_3^\text{3D} = J_{U(1)_b}^\text{3D} = 1/2$.

It was shown in
\cite{Chester:2021gdw} that when $m_2 = i\frac{b-b^{-1}}{2}$, the above
matrix model expression \eqref{eq:ZS3} reduces to a function only of
\begin{align}
	b^{-1}m_+~,\qquad bm_-~,
\end{align}
where $m_\pm \equiv m_3\pm m_1$. This means that, when
\eqref{eq:cond-Schur} is imposed, $Z_{S^3_b}$ depends only on
\begin{align}
	b^{-1}\left((\xi+1)M + (\xi-1)i\frac{b}{2}\right)~,\qquad   b\left((\xi-1)M + (\xi+1)i\frac{b}{2}\right)~.
\end{align}
Note that, for generic values of $\xi$ (including $\xi=0$), the $b$-dependence of
$Z_{S^3_b}$ cannot be removed by the rescale
\eqref{eq:rescale-Schur}, which is inconsistent with the general
discussion on the squashing independence.
For the
$S^3_b$ partition function of the ABJM theory to have the right
squashing independence, one must have
\begin{align}
	\xi  = -1~.
	\label{eq:xi}
\end{align}
Indeed, only for this value of $\xi$, the $b$-dependence of $Z_{S^3_b}$
can be removed by \eqref{eq:rescale-Schur}. Therefore we conclude that
the correct value of $\xi$ is $\xi=-1$.
Note that this implies a non-trivial mixing between the 4D $\mathfrak{u}(3)_R$ symmetry and the $\mathfrak{u}(1)_b$ flavor symmetry of the ABJM
theory!

\subsection{Consistency check with the other limits}

In the previous section, we have seen that there are two more special
values of the mass parameters,
\eqref{eq:Coulomb-m} and
\eqref{eq:M}, that give rise to squashing independence in three
dimensions. We here check if \eqref{eq:xi} is consistent with it.

Let us first consider \eqref{eq:Coulomb-m}, corresponding to $t=pq$ in
four dimensions.
In this case, our
identification \eqref{eq:dict} of mass parameters reduces to
\begin{align}
	m_1 = \xi M~,\qquad m_2 = -i\frac{b+b^{-1}}{2}~,\qquad m_3 = M~.
\end{align}
When $\xi= -1$, this implies
\begin{align}
	m_+ \equiv m_3 + m_1 = 0~.
	\label{eq:Coulomb3}
\end{align}
We now ask whether the squashing independence occurs when
the above conditions on the mass parameters are imposed.
Indeed, it was shown in appendix A.1 of \cite{Minahan:2021pfv} that
the $b$-dependence of $Z_{S^3_b}$ disappears in the case that $m_2 = -i(b+b^{-1})/2$
and $m_+=0$.\footnote{To be precise, the condition discussed in
\cite{Minahan:2021pfv} is $m_3 = i(b+b^{-1})/2$ and $m_1 - m_2=0$~. Since
$Z_{S^3_b}$ is invariant under $m_2 \leftrightarrow m_3$, this is
equivalent to $m_2 = i(b+b^{-1})/2$ and $m_1-m_3=0$. Using the
invariance of $Z_{S^3_b}$ under $m_2 \to -m_2$ and $m_3\to -m_3$~, this
leads to the squashing independence at \eqref{eq:Coulomb3}.} This is a
strong evidence for our identification $\xi = -1$.

Another limit we consider is \eqref{eq:M}, corresponding to
$a=\sqrt{t}/p$ in four dimensions.
In this case, we expect that the
$b$-dependence of $Z^3_b$ can be removed by replacing $m_2$ as
\begin{align}
	m_2 \to b^{-1}m_2 - i\frac{b-b^{-1}}{2}~.
	\label{eq:shift-N=3}
\end{align}
To see this is indeed the case, first note that our
identification \eqref{eq:dict} is now expressed as
\begin{align}
	m_1 = \xi\left(m_2 + ib\right)~,\qquad m_3 = i\frac{b-b^{-1}}{2}~.
	\label{eq:N=3-2}
\end{align}
It was shown in \cite{Chester:2021gdw} that, when $m_3=i(b-b^{-1})/2$, the partition function
$Z_{S^3_b}$ depends only on $b^{-1}(m_2+m_1)$ and
$b(m_2-m_1)$. Therefore, in
the case of $\xi = -1$, we see that $Z_{S^3_b}$ is a function of the single combination of variables
\begin{align}
	b\left(2 m_2 + ib\right)~.
\end{align}
It is now straightforward to show that the $b$-dependence of $Z_{S^3_b}$
is removed by \eqref{eq:shift-N=3} as expected. This is another strong
evidence for our identification $\xi=-1$.

\section{Divergence and flat directions}
\label{sec:divergence}

Having identified the correct R-symmetry mixing $\xi=-1$, we see that the mass parameters $m_1,m_2$ and $m_3$
of the ABJM theory are constrained by
\begin{align}
	m_1 = -\left(M+\frac{m}{2} + i\frac{b+b^{-1}}{4}\right)~,\qquad m_2 =
	m~, \qquad m_3 = M-\frac{m}{2}-i\frac{b+b^{-1}}{4}~,
	\label{eq:final-dict}
\end{align}
when it is obtained as the $S^1$-compactification of a 4D
$\mathcal{N}=3$ SCFT. Here, $M$ is related to the fugacity
$a$ for the $U(1)_f$ symmetry by $a=e^{-i\beta M}$, while $m$ is related to the
fugacity $t$ by $t=
	e^{-i\beta m -(b+b^{-1})/2}$.  The constraint $m_1 + m_2 + m_3 = -i(b+b^{-1})/2$ is interpreted to mean that the 3D global symmetry corresponding to
the mass parameter $(m_1 + m_2 + m_3)$ is accidental in three dimensions and has no counterpart in
four dimensions.

One important fact is that the $S^3_b$ partition function of the ABJM
theory is always divergent when \eqref{eq:final-dict} is imposed. Indeed, the
expression \eqref{eq:ZS3} for $Z_{S^3_b}$ contains the factor
\begin{align}
	\prod_{i,j=1}^N s_b\left(\frac{iQ}{4} -\left(\mu_i-\nu_j +
		\frac{m_1+m_2+m_3}{2}\right)\right) = \prod_{i,j=1}^N s_b\left(\frac{iQ}{2} -\mu_i +
	\nu_j\right)~,
	\label{eq:tw-hyper}
\end{align}
which is divergent whenever $\mu_i = \nu_j$. The divergence in
$Z_{S^3_b}$ is
schematically
\begin{align}
	\left(s_b\left(\frac{iQ}{2}\right)\right)^{N} =
	\left(\prod_{k,\ell=0}^\infty \frac{(k+1)b+(\ell+1)b^{-1}}{k b + \ell b^{-1}}\right)^N~.
	\label{eq:div-3d-ours}
\end{align}
Note that this divergence occurs even for the most general values of the
mass parameters, $m$ and $M$, that have 4D counterparts.

A similar divergence appears in the small $S^1$ limit of
the superconformal index of $\mathcal{N}=4$ super Yang-Mills
(SYM) theories \cite{ArabiArdehali:2015ybk}, as we will review
below in this paragraph. Recall that the
superconformal index is equivalent to the partition function of the
theory on $S^1\times S^3$ up to a prefactor.
For generic 4D $\mathcal{N}=2$ SCFTs, the small $S^1$ limit of the
$S^1\times S^3$ partition function $Z_{S^1\times S^3}$ behaves as
\cite{Imamura:2011uw, Spiridonov:2012ww, Aharony:2013dha,
	ArabiArdehali:2014otj, DiPietro:2014bca, ArabiArdehali:2015iow, Buican:2015ina}
\begin{align}
	\log Z_{S^1\times S^3} \sim \frac{8\pi^2}{\beta}(a-c) + \log Z_{S^3} +
	\mathcal{O}(\beta)~,
	\label{eq:asym}
\end{align}
where $a$ and $c$ are two conformal anomalies, and $Z_{S^3}$ is the
$S^3$ partition function of the 3D reduction of the 4D theory. However,
for $\mathcal{N}=4$ super Yang-Mills theories, the above formula is
modified as \cite{ArabiArdehali:2015ybk}
\begin{align}
	\log Z_{S^1\times S^3} \sim N\log \frac{2\pi}{\beta}+
	\mathcal{O}(\beta^0)~,
	\label{eq:asym-N=4}
\end{align}
where $N$ is the complex dimension of the ($\mathcal{N}=2$) Coulomb
branch. Note that the first term in \eqref{eq:asym} drops out since
$\mathcal{N}=4$ superconformal symmetry implies $a=c$. The divergent
term $N\log(2\pi/\beta)$ in \eqref{eq:asym-N=4} is interpreted to mean that $Z_{S^3}$ obtained in the small
$S^1$ limit of $Z_{S^1\times S^3}$ has a power-law
divergence as \cite{ArabiArdehali:2015ybk}
\begin{align}
	Z_{S^3} \sim \Lambda^N~,
	\label{eq:div-3d}
\end{align}
where $\Lambda$ is the cutoff for the vacuum expectation value (VEV) of the Coulomb branch
operators. The expression \eqref{eq:div-3d} implies that there are
$N$ flat directions in the 3D Coulomb branch on $S^3$ when the 3D theory
is obtained by an RG-flow from
four dimensions.\footnote{Here, the 3D ``Coulomb branch'' is a sub-space
	of the 3D moduli space on which the action of $\mathfrak{su}(2)_R$ is trivial, where
	$\mathfrak{su}(2)_L\times \mathfrak{su}(2)_R$ is an 3D $\mathcal{N}=4$
	R-symmetry.} Note that this divergence is purely three-dimensional;
$Z_{S^1\times S^3}$ is not divergent when the radius of $S^1$ is non-vanishing.

We now argue that our divergence in \eqref{eq:div-3d-ours} is precisely of
the form \eqref{eq:div-3d}. Indeed, each double-sine function
in \eqref{eq:tw-hyper} arises from the path integral of a
massless chiral multiplet whose VEV parameterizes a sub-space of the
Coulomb branch. Therefore, the zeros in the denominator of \eqref{eq:div-3d-ours}
imply that the 3D Coulomb branch has $N$ flat directions, leading to
the behavior \eqref{eq:div-3d} when the cutoff $\Lambda$ is introduced. The other directions on the Coulomb
branch are lifted by the background gauge fields and masses.\footnote{Note
	that the FI-parameter of the ABJM theory is already converted into a
	mass parameter via $\mathcal{N}=6$ supersymmetry. Therefore, the Coulomb
	branch can be lifted by changing the values of the mass parameters.}

Note that the above $N$ flat directions disappear when the
constraint $m_1 + m_2 + m_3 = -i(b+b^{-1})/2$ is relaxed.
In other words, this constraint on the mass parameters is such that the
$N$ directions on the Coulomb branch is unlifted. Here, $N$ is precisely the complex
dimension of the 4D Coulomb branch.
This suggest that, even for 4D
$\mathcal{N}=3$ SCFTs, the $\beta \to 0$ limit of the $S^1\times S^3$
partition function behaves as \eqref{eq:asym-N=4}.\footnote{Note that
	any 4D
	$\mathcal{N}=3$ SCFT has $a=c$ \cite{Aharony:2015oyb} and therefore the
	first term in the RHS of \eqref{eq:asym} vanishes for all 4D
	$\mathcal{N}\geq 3$ SCFTs.}

The above discussion implies that our identification $\xi=-1$ for the
R-symmetry mixing also resolves the second contradiction discussed in
Sec.~\ref{sec:intro}. To see this, let us focus on the 4D rank-one
$\mathcal{N}=3$ SCFTs by setting $N=1$. The 4D moduli
space is then $\mathbb{C}^3/\mathbb{Z}_k$ for $k=3,4$ or $6$, and the 3D moduli space
is $\mathbb{C}^4/\mathbb{Z}_k$. We denote their local coordinates by
$(z_1,z_2,z_3)$ and $(z_1,z_2,z_3,z_4)$, respectively. The extra coordinate $z_4$
of the 3D moduli space corresponds to the VEV of the scalar field in the
massless chiral multiplet discussed above. The fact that the $z_4$-direction is flat
(even with the most general mass parameters arising from four dimensions turned
on) implies that, under our identification of the R-charges
\eqref{eq:4D-3D} with $\xi=-1$, the coordinate $z_4$
is neutral under
$\mathfrak{u}(3)_R$ as expected. This resolves the second contradiction.
Note that this neutrality of $z_4$ is
possible only when there is a non-trivial mixing between
the $\mathfrak{u}(1)_c$ and $\mathfrak{u}(1)_b$.

Finally, we comment that there is an extra divergence in
$Z_{S^3_b}$ when we set $m_2 = m = -i(b+b^{-1})/2$. Indeed, in this case, we
have $m_1+m_3=0$ and therefore
\begin{align}
	\prod_{i,j=1}^{N}s_b\left(\frac{iQ}{4}-\left(-\mu_i+\nu_j+ \frac{-m_1+m_2-m_3}{2}\right)\right)
\end{align}
has exactly the same divergence as \eqref{eq:tw-hyper}. The reason for
this extra divergence is that, when $m =
	-i(b+b^{-1})/2$, the 4D partition function $Z_{S^1\times S^3}$ is already divergent even before
taking the small $S^1$ limit. To see this, recall that $m=-i(b+b^{-1})/2$ corresponds
to $t=pq$ in the 4D index, as discussed in
Sec.~\ref{subsec:review}. Since a 4D Coulomb branch operator
of $U(1)_r$ charge $r$ contributes $(t/pq)^r$ to the index,\footnote{It
	is known that every Coulomb branch operator in any 4D $\mathcal{N}\geq
		2$ SCFT is neutral under $\mathcal{N}=2$ flavor symmetries
	\cite{Buican:2013ica}. See also \cite{Buican:2014qla} for its higher spin generalization.} the
condition $t=pq$ sets all the index contributions from an infinite
number of Coulomb
branch operators to $1$, leading to a divergence. This reflects the fact that the 4D Coulomb branch is not lifted
when $t=pq$. This is in contrast to the case of generic values
of $m$, in which case the 4D index is
finite but the 3D partition function $Z_{S^3}$ is divergent.

\section{Conclusions and Discussions}

\label{sec:conclusions}

In this paper, we have studied the $S^1$-compactification of 4D
$\mathcal{N}=3$ SCFTs, focusing on the relation between the 4D
superconformal index and 3D partition function $Z_{S^3_b}$. In particular,
we have argued that the center $\mathfrak{u}(1)$ of $\mathfrak{u}(3)_R$
in four dimensions can mix with an abelian flavor symmetry of the 3D
$\mathcal{N}=6$ theory obtained by the compactification. In the case that the 3D theory is
the ABJM theory, we have shown that such an R-symmetry mixing does occur
and is
uniquely fixed so that the Schur limit (and/or its $\mathcal{N}=3$
cousin) of the 4D index correctly reproduces the squashing independence
of $Z_{S^3_b}$ found in \cite{Chester:2021gdw}. Our result implies that the
recent discussions in \cite{Minahan:2021pfv} on the connection between supersymmetry enhancement
of the 4D index and squashing independence of
$Z_{S^3_b}$ can also be applied to the ABJM theories.

The R-symmetry mixing we have found is also consistent
with the expectation that the 4D $\mathfrak{u}(3)_R$ trivially acts on
the sub-space of the 3D Coulomb branch that does not have a 4D
counterpart (and therefore purely three-dimensional). This trivial
action of $\mathfrak{u}(3)_R$ implies that the 3D limit of 4D
$\mathcal{N}=3$ index
leads to a divergence. A similar
divergence appears in the case of 4D $\mathcal{N}=4$ SYM theories. This
seems to suggest that the small $S^1$ limit of the superconformal index of 4D
$\mathcal{N}\geq 3$ SCFTs always behaves as \eqref{eq:asym-N=4}.

We note that the R-symmetry mixing that we have discussed in this paper is
similar (but different) to the one for the compactification of
$\mathcal{N}=2$ Argyres-Douglas (AD) SCFTs \cite{Buican:2015hsa}. When a generic AD
theory is compactified on $S^1$, there is a mixing between the 4D $\mathfrak{u}(1)_r$
symmetry and 3D topological $\mathfrak{u}(1)$ global symmetry. This is because the
4D $\mathcal{N}=2$ R-symmetry, $\mathfrak{su}(2)_R \times \mathfrak{u}(1)_r$, needs to
enhance to the 3D $\mathcal{N}=4$ R-symmetry, $\mathfrak{so}(4)_R \simeq \mathfrak{su}(2)_R\times
	\mathfrak{su}(2)_C$, under the compactification; without the mixing between
$\mathfrak{u}(1)_r$ and the topological $\mathfrak{u}(1)$, such an
enhancement is prohibited by the presence of Coulomb branch operators of
fractional $\mathfrak{u}(1)_r$ charges. The R-symmetry mixing we
have discussed in this paper is, in contrast, the mixing between the center
of the 4D $\mathcal{N}=3$ R-symmetry,
$\mathfrak{u}(3)_R$, and the 3D $\mathfrak{u}(1)_b$
symmetry. Since $\mathfrak{u}(1)_b$ can be regarded as a topological
$\mathfrak{u}(1)$ symmetry of the Chern-Simons matter theory, our
R-symmetry mixing shares some characteristics with the one for the AD
theories. A big difference is, however, 4D $\mathcal{N}=3$ SCFTs have no
Coulomb branch operators of fractional $\mathfrak{u}(1)_r$ charges. The
R-symmetry mixing in our case just reflects the $\mathcal{N}=3$ superconformal symmetry.

In general, our
       result gives a necessary condition for a 3D
       $\mathcal{N}=6$ SCFT to have its 4D uplift. Indeed, whenever a 3D
       $\mathcal{N}=6$ SCFT can be realized by compactifying a 4D
       $\mathcal{N}=3$ SCFT, there must be special sub-spaces in the
       space of 3D mass parameters in which $Z_{S^3}$ is
       squashing independent. These sub-spaces correspond to the 4D
       Schur limit $t=q$ and its $\mathcal{N}=3$ cousins
       $a=\sqrt{t}/p$.\footnote{The 3D partition function $Z_{S^3}$ must also be squashing
       independent at $t=pq$, when the 3D theory has a 4D uplift.} Therefore, if the space
       of mass parameters of a 3D $\mathcal{N}=6$ SCFT has no such
       special sub-spaces, one can conclude that there exists no 4D
       $\mathcal{N}=3$ SCFT whose $S^1$-compactification gives rise to that 3D $\mathcal{N}=6$
       SCFT.

There are clearly many future directions, some of which we list below:
\begin{itemize}
	\item While we have focused on 4D $\mathcal{N}=3$ SCFTs whose 3D
	      reduction is the ABJM theory, there are many other
	      $\mathcal{N}=6$ Chern-Simons matter theories
	      \cite{Hosomichi:2008jb, Aharony:2008gk}. It would
	      be interesting to identify the R-symmetry mixing for these other
	      cases. In
	      particular, the generalization to the ABJ theory must be
	      straightforward. 

	\item It is known that the stress-tensor multiplet of every 3D
	      $\mathcal{N}=6$ SCFT contains an $\mathcal{N}=6$ flavor current \cite{Bashkirov:2011fr}
	      (See also Sec.~5.4.6 of \cite{Cordova:2016emh}). In the case of ABJM theories, this is
	      the $\mathfrak{u}(1)_b$ current. Based on our discussion, it is
	      natural to
	      expect that there is a mixing between the 4D $\mathfrak{u}(3)_R$ current
	      and this $\mathcal{N}=6$ flavor current. It would be interesting
	      to see if there is a universal formula for this R-symmetry mixing.

	\item It is desirable to obtain a closed form expression for the
	      superconformal index (or at least its Schur limit) of a 4D
	      $\mathcal{N}=3$ SCFT. Several limits of the superconformal index of 4D $\mathcal{N}=3$
	      SCFTs are computed in \cite{Imamura:2016abe, Bonetti:2018fqz, Arai:2019xmp, Zafrir:2020epd,
		      Agarwal:2021oyl}, but a closed form expression for the index with
	      a non-trivial flavor
	      fugacity is not known. Once we find such an
	      expression, we should be able to study its small $S^1$
	      limit. Then it
	      would be interesting to see if this small $S^1$ limit is
	      identical (up to a prefactor) to $Z_{S^3_b}$ of the 3D $\mathcal{N}=6$
	      theory. For this purpose, one needs to take into account the
	      R-symmetry mixing that we discussed in this paper.
\end{itemize}

\ack{We are grateful to M.~Buican, Y.~Hatsuda, Y.~Nakayama,
	Y.~Tachikawa and Y.~Yoshida for helpful discussions.
	T.~Nakanishi’s research is partially supported by JST, the establishment of university fellowships towards the creation of science technology innovation,  Grant Number JPMJFS2138.
	T.~Nishinaka’s research is partially supported by JSPS KAKENHI Grant Numbers JP18K13547 and JP21H04993.
	This work was also partially supported by Osaka Central Advanced Mathematical Institute: MEXT Joint Usage/Research Center on Mathematics and Theoretical Physics JP- MXP0619217849.
}

\newpage

\begin{appendices}

	\section{4D $\mathcal{N}=3$ and 3D $\mathcal{N}=6$ superconformal algebras}
	\label{app:R-symmetry}

	In this section, we discuss in some more detail how the 4D
	and 3D symmetries are related under the $S^1$-compactification.

	\subsection{4D $\mathcal{N}=3$ superconformal algebra}

	\label{app:4Dsca}

	In this subsection, we review the 4D $\mathcal{N}=3$ superconformal algebra \cite{Dolan:2002zh}.
	We follow the convention of Appendix A of
	\cite{Beem:2013sza} unless otherwise stated.
	We use $\sigma^\mu=(I_2,\sigma^1,\sigma^3,\sigma^2)$ and
	$\bar{\sigma}^\mu=(-I_2,\sigma^1,\sigma^3,\sigma^2)$ with
	$I_2$ being the $2\times 2$ identity matrix.\footnote{Note that this
		means $\bar{\sigma}^{\mu\dot\alpha\alpha} = \epsilon^{\alpha
				\beta}\sigma^\mu_{\beta\dot\beta}\epsilon^{\dot\beta \dot\alpha}$.}

	The 4D $\mathcal{N}=3$ superconformal algebra (for Minkowski signature)
	is $\mathfrak{su}(2,2|
		3)$, which contains $\mathfrak{su}(2,2)\oplus\mathfrak{u}(3)_R$ as the
	bosonic sub-algebra.
	We take a basis of $\mathfrak{su}(2,2|3)$ as
	\begin{gather}
		\overbrace{\mathcal{M}_{\alpha}\!^{\beta}~,\quad
		\tilde{\mathcal{M}}^{\dot{\alpha}}\!_{\dot{\beta}}~,\quad
		\mathcal{P}_{\alpha\dot{\beta}}~,\quad
		\mathcal{K}^{\dot{\alpha}\beta}~,\quad
		\mathcal{H}
		}^{\mathfrak{su}(2,2)}~,\quad
		\overbrace{\mathcal{R}^I\!_J~}^{\mathfrak{u}(3)_R},\quad
		\mathcal{Q}^I\!_{\alpha}~,\quad
		\tilde{\mathcal{Q}}_{I\dot{\alpha}}~,\quad
		\mathcal{S}_I\!^{\alpha}~,\quad
		\tilde{\mathcal{S}}^{I\dot{\alpha}}~,
		\label{eq:4dMgen2}
	\end{gather}
	where $I,J=1,2,3$, and the first
	six are bosonic while the last four are fermionic.
	The non-vanishing commutation relations for the bosonic charges are
	written as
	\begin{align}
		[\mathcal{M}_\alpha{}^\beta,\,\mathcal{M}_\gamma{}^\delta] =
		\delta_\gamma^\beta \mathcal{M}_{\alpha}{}^\delta -
		\delta_{\alpha}^\delta\mathcal{M}_\gamma{}^\beta~                                & ,\qquad
		[\tilde{\mathcal{M}}^{\dot\alpha}{}_{\dot\beta},\,
			\tilde{\mathcal{M}}^{\dot\gamma}{}_{\dot\delta}] =
		\delta^{\dot\alpha}_{\dot\delta}\tilde{\mathcal{M}}^{\dot\gamma}{}_{\dot\beta}
		-
		\delta^{\dot\gamma}_{\dot\beta}\tilde{\mathcal{M}}^{\dot\alpha}{}_{\dot\delta}~,
		\label{eq:4D_MM}
		\\
		[\mathcal{M}_\alpha{}^\beta,\, \mathcal{P}_{\gamma\dot\gamma}] = \mathcal{P}_{\alpha\dot\gamma}\delta^\beta_\gamma -
		\frac{1}{2}\delta^\beta_\alpha\mathcal{P}_{\gamma\dot\gamma}~                    & ,\qquad
		[\tilde{\mathcal{M}}^{\dot\alpha}{}_{\dot\beta},\,\mathcal{P}_{\gamma\dot\gamma}]
		= \delta^{\dot\alpha}_{\dot\gamma}\mathcal{P}_{\gamma\dot\beta} -
		\frac{1}{2}\delta^{\dot\alpha}_{\dot\beta}\mathcal{P}_{\gamma\dot\gamma}~,
		\label{eq:4D_MP}
		\\
		[\mathcal{M}_\alpha{}^\beta,\, \mathcal{K}^{\dot\gamma\gamma}] = -\mathcal{K}^{\dot\gamma \beta}\delta^\gamma_\alpha+
		\frac{1}{2}\delta^\beta_\alpha\mathcal{K}^{\dot\gamma\gamma}~                    & ,\qquad
		[\tilde{\mathcal{M}}^{\dot\alpha}{}_{\dot\beta},\,\mathcal{K}^{\dot\gamma\gamma}]
		= -\mathcal{K}^{\dot\alpha\gamma}\delta^{\dot\gamma}_{\dot\beta}
		+
		\frac{1}{2}\delta^{\dot\alpha}_{\dot\beta}\mathcal{K}^{\dot\gamma\gamma}~,
		\\
		[\mathcal{K}^{\dot\alpha\alpha},\,\mathcal{P}_{\beta\dot\beta}]                  & =
		\frac{1}{4}(\delta^{\dot\alpha}_{\dot\beta}\mathcal{M}_{\beta}{}^{\alpha}
		+ \delta^{\alpha}_\beta\tilde{\mathcal{M}}^{\dot\alpha}{}_{\dot\beta}+
		\delta^{\alpha}_{\beta}\delta^{\dot\alpha}_{\dot\beta}\mathcal{H})~,
		\\
		[\mathcal{H},\,\mathcal{P}_{\alpha\dot\alpha}] = \mathcal{P}_{\alpha\dot\alpha}~ & ,\qquad
		[\mathcal{H},\,\mathcal{K}^{\dot\alpha\alpha}] = -\mathcal{K}^{\dot\alpha\alpha}~,
		\\
		\qquad\quad[\mathcal{R}^I\!_J,\mathcal{R}^K\!_L]
		=                                                                                & \delta^K_J \mathcal{R}^I\!_L-\delta^I_L \mathcal{R}^K\!_J~.
		\label{eq:4D_RR}
	\end{align}
	Our notation for the R-charges is such that
	\begin{align}
		[\mathcal{R}^I\!_J,\mathcal{Q}^K_\alpha]
		=\delta^K_J\mathcal{Q}^I\!_\alpha-\frac{1}{4}\delta^I_J\mathcal{Q}^K\!_\alpha~,                           & \quad
		[\mathcal{R}^I\!_J,\tilde{\mathcal{Q}}_{K\dot{\alpha}}]
		=-\delta^I_K\tilde{\mathcal{Q}}_{J\dot{\alpha}}+\frac{1}{4}\delta^I_J\tilde{\mathcal{Q}}_{K\dot{\alpha}}~,
		\label{eq:4D_RQ}
		\\
		[\mathcal{R}^I\!_J,\tilde{\mathcal{S}}^{K\dot{\alpha}}]
		=\delta^K_J\tilde{\mathcal{S}}^{I\dot{\alpha}}-\frac{1}{4}\delta^I_J\tilde{\mathcal{S}}^{K\dot{\alpha}}~, & \quad
		[\mathcal{R}^I\!_J,\mathcal{S}_K\!^\alpha]
		=-\delta^I_K\mathcal{S}_J\!^\alpha+\frac{1}{4}\delta^I_J\mathcal{S}_K\!^\alpha~.
		\label{eq:4dMRcom}
	\end{align}
	The non-trivial anti-commutation relations for the supercharges are
	written as
	\begin{align}
		\{\mathcal{Q}^I_\alpha,\tilde{\mathcal{Q}}_{J\dot{\alpha}}\}
		=2\delta^I_J
		\mathcal{P}_{\alpha\dot\alpha}
		~                               & ,\quad
		\{\mathcal{Q}^I_\alpha,\mathcal{S}_J^\beta\}
		=\frac{1}{2}\delta^I_J\delta^\beta_\alpha\mathcal{H}+\delta^I_J\mathcal{M}_\alpha\!^\beta-\delta^\beta_\alpha\mathcal{R}^I\!_J~,
		\label{eq:4D_QQ}
		\\[1mm]
		\{\tilde{\mathcal{S}}^{I\dot{\alpha}},\mathcal{S}_J^{\alpha}\}
		=2\delta^I_J
		\mathcal{K}^{\dot\alpha\alpha}~ & ,\quad
		\{\tilde{\mathcal{S}}^{I\dot{\alpha}},\tilde{\mathcal{Q}}_{J\dot{\beta}}\}
		=\frac{1}{2}\delta^I_J\delta^{\dot{\alpha}}_{\dot{\beta}}\mathcal{H}+\delta^I_J\tilde{\mathcal{M}}^{\dot{\alpha}}\!_{\dot{\beta}}+\delta^{\dot{\alpha}}_{\dot{\beta}}\mathcal{R}^I\!_J~.
	\end{align}
	Here, our normalization of $\mathcal{P}_{\alpha\dot\alpha}$ and
	$\mathcal{K}^{\dot\alpha\alpha}$ are different from that in
	\cite{Beem:2013sza} by a factor two.
	The remaining non-vanishing commutation relations are as follows:
	\begin{align}
		[\mathcal{H},\mathcal{Q}^I\!_\alpha]
		=\frac{1}{2}\mathcal{Q}^I\!_\alpha~,\quad
		[\mathcal{H},\tilde{\mathcal{Q}}_{I\dot{\alpha}}]
		=\frac{1}{2}\tilde{\mathcal{Q}}_{I\dot{\alpha}}~                                                                                                      & ,\quad
		[\mathcal{H},\mathcal{S}_I\!^\alpha]
		=-\frac{1}{2}\mathcal{S}_I\!^\alpha~,\quad
		[\mathcal{H},\tilde{\mathcal{S}}^{I\dot{\alpha}}]
		=-\frac{1}{2}\tilde{\mathcal{S}}^{I\dot{\alpha}}~,
		\\
		[\mathcal{M}_\alpha\!^\beta,\mathcal{Q}^I\!_\gamma]
		=\delta^\beta_\gamma\mathcal{Q}^I\!_\alpha-\frac{1}{2}\delta^\beta_\alpha\mathcal{Q}^I\!_\gamma~                                                      & ,\quad
		[\mathcal{M}_\alpha\!^\beta,\mathcal{S}_I\!^\gamma]
		=-\delta^\gamma_\alpha\mathcal{S}_I\!^\beta+\frac{1}{2}\delta^\beta_\alpha\mathcal{S}_I\!^\gamma~,
		\label{eq:4D_MQ1}
		\\
		[\tilde{\mathcal{M}}^{\dot{\alpha}}\!_{\dot{\beta}},\tilde{\mathcal{Q}}_{I\dot{\gamma}}]
		=\delta^{\dot{\alpha}}_{\dot{\gamma}}\tilde{\mathcal{Q}}_{I\dot{\beta}}-\frac{1}{2}\delta^{\dot{\alpha}}_{\dot{\beta}}\tilde{\mathcal{Q}}^I\!_\gamma~ & ,\quad
		[\tilde{\mathcal{M}}^{\dot{\alpha}}\!_{\dot{\beta}},\tilde{\mathcal{S}}^{I\dot{\gamma}}]
		=-\delta^{\dot{\gamma}}_{\dot{\beta}}\tilde{\mathcal{S}}^{I\dot{\alpha}}+\frac{1}{2}\delta^{\dot{\alpha}}_{\dot{\beta}}\tilde{\mathcal{S}}^{I\dot{\gamma}}~,
		\label{eq:4D_MQ2}
		\\
		[\mathcal{K}^{\dot{\alpha}\alpha},\mathcal{Q}^I\!_\beta]
		=\frac{1}{2}\delta^\alpha_\beta\tilde{\mathcal{S}}^{I\dot{\alpha}}~                                                                                   & ,\quad
		[\mathcal{K}^{\dot{\alpha}\alpha},\tilde{\mathcal{Q}}_{I\dot{\beta}}]
		=\frac{1}{2}\delta^{\dot{\alpha}}_{\dot{\beta}}\mathcal{S}_I\!^\alpha~,                                                                                        \\
		[\mathcal{P}_{\alpha\dot{\alpha}},\mathcal{S}_I^\beta]
		=-\frac{1}{2}\delta^\beta_\alpha\tilde{\mathcal{Q}}_{I\dot{\alpha}}~                                                                                  & ,\quad
		[\mathcal{P}_{\alpha\dot{\alpha}},\tilde{\mathcal{S}}^{I\dot{\beta}}]
		=-\frac{1}{2}\delta^{\dot{\beta}}_{\dot{\alpha}}\mathcal{Q}^I\!_\alpha~.
	\end{align}

	The hermiticity is such that
	\begin{align}
		\mathcal{H}^\dagger            & = \mathcal{H}~,\qquad
		(\mathcal{P}_{\alpha\dot\beta})^\dagger =
		\mathcal{K}^{\dot\beta\alpha}
		~,\qquad
		(\mathcal{M}_\alpha{}^\beta)^\dagger =
		\mathcal{M}_\beta{}^\alpha~,\qquad
		(\tilde{\mathcal{M}}^{\dot\alpha}{}_{\dot\beta})^\dagger =
		\tilde{\mathcal{M}}^{\dot\beta}{}_{\dot\alpha}~,
		\\
		(\mathcal{Q}^I_\alpha)^\dagger & =
		\mathcal{S}_I^\alpha~,\qquad
		(\tilde{\mathcal{Q}}_{I\dot\alpha})^\dagger =
		\tilde{\mathcal{S}}^{I\dot\alpha}~,\qquad (\mathcal{R}^I{}_J)^\dagger =
		\mathcal{R}^J{}_I~.
	\end{align}

	\subsection{3D $\mathcal{N}=6$ superconformal algebra}

	We here review the 3D $\mathcal{N}=6$ superconformal algebra, following
	Appendix B of \cite{Chester:2014mea}. We use the 3D gamma matrices
	$\gamma^i_{\alpha\beta} = (I_2,\,\sigma^1,\,\sigma^3)_{\alpha\beta}$ and
	$\bar{\gamma}^{i\alpha\beta} = (-I_2,\,\sigma^1,\,\sigma^3)$.

	The 3D $\mathcal{N}=6$ superconformal algebra is $\mathfrak{osp}(6|4)$
	whose basis we take as
	\begin{align}
		\overbrace{M_{\alpha}{}^{\beta}~,\quad
			P_{\alpha\beta}~,\quad
			K^{\alpha\beta}~,\quad
			D
		}^{\mathfrak{so}(3,2)}~,\quad
		\overbrace{R_{rs}~}^{\mathfrak{so}(6)_R},\quad
		Q_{\alpha r}~,\quad
		S^{\alpha}{}_r~,
	\end{align}
	where $P_{\alpha\beta}$ and $K^{\alpha\beta}$ are symmetric under
	$\alpha\leftrightarrow \beta$ while $R_{st}$ is anti-symmetric under
	$s\leftrightarrow t$.
	They satisfy the following (anti-)commutation relations:
	\begin{align}
		[M_\alpha{}^\beta,\,M_{\gamma}{}^\delta]                       & = -\delta_\alpha^\delta
		M_\gamma{}^\beta + \delta^\beta_\gamma M_\alpha{}^\delta~,
		\label{eq:3D_MM}
		\\
		[M_\alpha{}^\beta,\, P_{\gamma \delta}]                        & = \delta^\beta_\gamma
		P_{\alpha\delta} + \delta^{\beta}_{\delta} P_{\alpha\gamma} -
		\delta^{\beta}_{\alpha}P_{\gamma\delta}~,
		\label{eq:3D_MP}
		\\
		[M_\alpha{}^\beta,\, K^{\gamma\delta}]                         & = -\delta^\gamma_\alpha
		K^{\beta\delta} - \delta^{\delta}_{\alpha} K^{\beta\gamma} +
		\delta^{\beta}_\alpha K^{\gamma\delta}~,
		\\
		[K^{\alpha\beta},\,P_{\gamma\delta}]                           & = 4
		\delta^{(\alpha}_{(\gamma}M_{\delta)}{}^{\beta)} + 4
		\delta^{(\alpha}_{(\gamma}\delta^{\beta)}_{\delta)}D~,
		\\
		[D,\,P_{\alpha\beta}] = P_{\alpha\beta}~                       & ,\qquad [D,\,K^{\alpha\beta}] =
		-K^{\alpha\beta}~,
		\\
		[R_{rs},\,R_{tu}]                                              & =i\left(\delta_{rt} R_{su} - \delta_{st}R_{ru}
		-\delta_{ru}R_{st} + \delta_{su}R_{rt}\right)~,
		\label{eq:3D_RR}
		\\
		[R_{rs},\,Q_{\alpha t}]                                        & = i(\delta_{rt}Q_{\alpha s} -
		\delta_{st}Q_{\alpha r})~,
		\label{eq:3D_RQ}
		\\
		[R_{rs},\,S^\alpha{}_t]                                        & = i\left(\delta_{rt} S^\alpha{}_s - \delta_{st}S^\alpha{}_r\right)
		\\
		\{Q_{\alpha r},\, Q_{\beta s}\} = 2\delta_{rs}P_{\alpha\beta}~ & ,\qquad
		\{S^\alpha{}_r,\,S^\beta{}_s\} = -2\delta_{rs}K^{\alpha\beta}~,
		\label{eq:3D_QQ}
		\\
		\{Q_{\alpha r},\, S^\beta{}_{s}\}
		                                                               & =2i\left(\delta_{rs}\delta^\beta_\alpha D +
		\delta_{rs}M_\alpha{}^\beta -i \delta^\beta_\alpha R_{rs}\right)
		\\
		[D,Q_{\alpha r}] = \frac{1}{2}Q_{\alpha r}~                    & ,\qquad [D,S^{\alpha}{}_r]
		=-\frac{1}{2}S^\alpha{}_r~,
		\\
		[M_\alpha{}^\beta,\,Q_{\gamma r}]                              & = \delta^\beta_\gamma Q_{\alpha r} -
		\frac{1}{2}\delta^\alpha_\beta Q_{\gamma r}~,
		\label{eq:3D_MQ}
		\\
		[M_\alpha{}^\beta,\,S^\gamma{}_r]                              & = -\delta^\gamma_\alpha S^{\beta}{}_r
		+ \frac{1}{2}\delta^\beta_\alpha S^{\gamma}{}_r~,
		\\
		[K^{\alpha\beta},\, Q_{\gamma r}] = -2i \delta^{(\alpha}_\gamma
		S^{\beta)}{}_r~                                                & ,\qquad [P_{\alpha\beta},\, S^{\gamma}{}_r] = -2i
		\delta^{\gamma}_{(\alpha}Q_{\beta) r}
	\end{align}
	The hermiticity is expressed as
	\begin{align}
		D^\dagger =D~,\qquad (P_{\alpha\beta})^\dagger = K^{\alpha\beta}~,\qquad
		(M_\alpha{}^\beta)^\dagger = M_{\beta}{}^\alpha~,\qquad  (Q_{\alpha r})^\dagger = -iS^\alpha{}_r~,\qquad (R_{rs})^\dagger = R_{rs}~.
	\end{align}

	\subsection{Reducting 4D $\mathcal{N}=3$ to 3D $\mathcal{N}=6$}

	Let us consider the $S^1$-compactification of a 4D $\mathcal{N}=3$
	SCFT, which leads to a 3D $\mathcal{N}=6$ SCFT in the deep infrared. We
	here discuss how the 4D $\mathcal{N}=3$ supersymmetry is related to the
	3D $\mathcal{N}=6$ supersymmetry.

	Note that the $S^1$-compactification
	explicitly breaks the conformal symmetry. We therefore focus on the super Poincar\'e
	sub-algebra generated by
	$\mathcal{Q}_\alpha^I,\,\tilde{\mathcal{Q}}_{I\dot\alpha},\, \mathcal{R}^I{}_J,\,\mathcal{M}_\alpha{}^\beta,\,\tilde{\mathcal{M}}^{\dot\alpha}{}_{\dot\alpha}$
	and $\mathcal{P}_{\alpha\dot\beta}$. The relevant (anti-)commutation
	relations are Eqs.~\eqref{eq:4D_MM},\,
	\eqref{eq:4D_MP},\,\eqref{eq:4D_RR},\,\eqref{eq:4D_RQ} and the left
	relations of \eqref{eq:4D_QQ}, \eqref{eq:4D_MQ1} and \eqref{eq:4D_MQ2} . Below, we will identify how these
	relations are mapped to the (anti-)commutation relations for the 3D
	$\mathcal{N}=6$ super Poincar\'e algebra.

	To that end, we first identify the 3D supercharges as
	\begin{align}
		Q_{\alpha r} = \left\{
		\begin{array}{l}
			\frac{1}{\sqrt{2}}\left(\mathcal{Q}^I_\alpha + \tilde{\mathcal{Q}}_{I
				\alpha}\right)\quad \text{when} \;\; r=2I-1 \;\;\text{ for
			} \;\; I \in \{1,2,3\}
			\\
			\frac{i}{\sqrt{2}}\left(\mathcal{Q}^I_\alpha - \tilde{\mathcal{Q}}_{I
				\alpha}\right)\quad \text{when} \;\; r=2I \;\;\text{ for
			} \;\; I \in \{1,2,3\}
			\\
		\end{array}
		\right.~.
	\end{align}
	Note that the 3D Lorentz symmetry $\mathfrak{so}(1,2)\simeq \mathfrak{sl}(2,\mathbb{R})$ is the diagonal sub-algebra of the 4D Lorentz
	symmetry $\mathfrak{so}(1,3)\simeq \mathfrak{sl}(2,\mathbb{C})\simeq \mathfrak{sl}(2,\mathbb{R})^2$, and therefore we
	identify $\dot\alpha$ and $\alpha$ under the compactification. This
	means that the 3D Lorentz generator is identified as
	\begin{align}
		M_\alpha{}^\beta = \mathcal{M}_\alpha{}^\beta + \tilde{\mathcal{M}}^{\beta}{}_\alpha~.
	\end{align}
	The three-dimensional translations are
	\begin{align}
		P_{\alpha\beta} = \frac{1}{2}\left(\mathcal{P}_{\alpha \beta} + \mathcal{P}_{\beta\alpha}\right)~,
	\end{align}
	which removes the ``fourth'' component of the momentum.
	One can check that these identifications correctly reproduce the 3D
	(anti-)commutation relations
	\eqref{eq:3D_MM},\,\eqref{eq:3D_MP},\,
	\eqref{eq:3D_MQ} and the first relation in \eqref{eq:3D_QQ}.\footnote{In the reduction to three dimensions, we set
		$\mathcal{P}_{\alpha\beta}-\mathcal{P}_{\beta\alpha} = 0$ since we only
		keep the zero modes in the compactified direction.}

	We now turn to the R-symmetry. Recall that the 4D R-symmetry is
	$\mathfrak{u}(3)_R$ while the 3D R-symmetry is
	$\mathfrak{so}(6)_R$. Along the RG-flow from four dimensions
	to three dimensions, the manifest global symmetry is only
	$\mathfrak{u}(3)_R$, which is in the deep infrared identified with a sub-algebra
	of the product of $\mathfrak{so}(6)_R$ and the $\mathcal{N}=6$ flavor
	symmetry. In the main text, we argue that the center $\mathfrak{u}(1)$
	of $\mathfrak{u}(3)_R$ is mixed with the $\mathcal{N}=6$ flavor
	symmetry. This is generally expressed (up to automorphims) as
	\begin{align}
		\mathcal{R}^1{}_1 & = -\frac{3}{4}R_{12} + \frac{1}{4}R_{34} +
		\frac{1}{4}R_{56} + \frac{\xi}{4} J_\text{flavor}^\text{3D}~,
		\\
		\mathcal{R}^2{}_2 & = +\frac{1}{4}R_{12} - \frac{3}{4}R_{34} +
		\frac{1}{4}R_{56} + \frac{\xi}{4} J^\text{3D}_\text{flavor}~,
		\\
		\mathcal{R}^3{}_3 & = +\frac{1}{4}R_{12} + \frac{1}{4}R_{34}
		-\frac{3}{4}R_{56} + \frac{\xi}{4} J^\text{3D}_\text{flavor}~,
	\end{align}
	where $J^\text{3D}_\text{flavor}$ is an $\mathcal{N}=6$ flavor symmetry
	that mixes with the center of $\mathfrak{u}(3)_R$ and $\xi$ is a real
	parameter. When the 3D theory is the ABJM theory,
	$J^\text{3D}_\text{flavor}$ is the $\mathfrak{u}(1)_b$
	charge.\footnote{In comparison to Eq.~\eqref{eq:4D-3D}, we identify
	$J^{\text{3D}}_{\text{flavor}} = \frac{1}{4}J_{U(1)_b}^{\text{3D}}$.}and
	$\xi=-1$ for the normalization that we used in the main text.

	One can also identify how the non-abelian part of the 4D R-symmetry, $\mathfrak{su}(3)\subset
		\mathfrak{u}(3)_R$, is embedded in the 3D global symmetry. Indeed, under
	the $S^1$-compactification, the 4D
	R-charges $\mathcal{R}^I{}_J$ for $I\neq J$ are identified as the following linear
	combinations of the 3D R-charges $R_{rs}$:\footnote{One can summarize
	these six relations in $\mathcal{R}^I{}_J = \frac{1}{2}\left(R_{2J\,
			2I-1} + R_{2I\, 2J-1} - iR_{2J-1\,2I-1} -iR_{2J\,2I}\right)$. Note that
	$R_{rs} = -R_{sr}$.}
	\begin{align}
		\mathcal{R}^1{}_2 & = \frac{1}{2}(-R_{14}+R_{23}+iR_{13}+iR_{24})~,\qquad
		\mathcal{R}^2{}_1 = \frac{1}{2}(-R_{14}+R_{23} -iR_{13}-iR_{24})~,
		\\
		\mathcal{R}^2{}_3 & = \frac{1}{2}(-R_{36}+R_{45}+iR_{35} +
		iR_{46})~,\qquad \mathcal{R}^3{}_2 = \frac{1}{2}(-R_{36}+R_{45}-iR_{35}
		-iR_{46})~,
		\\
		\mathcal{R}^3{}_1 & = \frac{1}{2}(-R_{16}+R_{25}-iR_{15}-iR_{26})~,\qquad
		\mathcal{R}^1{}_3 = \frac{1}{2}(-R_{16}+R_{25}+iR_{15}+iR_{26})~.
	\end{align}
	It is straightforward to see that the above identifications are consistent
	with the 3D and 4D commutation relations for the R-charges and supercharges.

\end{appendices}

\bibliography{squashing}
\bibliographystyle{utphys}

\end{document}